# Making sense of Rule 605

Peter B. Lerner[1]

Abstract

The paper analyzes monthly reports of two companies: one market maker, Virtu, and one broker-dealer, SoHo, in the period between January 2021 and March 2024. The predictive power of standard methods for the probability laws of stock orders- linear regression and LSTM- is poor, despite their probability laws being relatively stable. However, the month-to-month correlations are highly stable and predictable. In this paper, we study the distributions of the sixteen datasets resulting from the 26 stock-trading parameters reported by brokerages under regulatory requirements.

## 1. Introduction

A general trend in financial market regulation is the desire of regulators and investors to disclose ever more data (Zhao 2007). In recent years, an increasing share of trading goes to dark venues through alternative and non-alternative trading systems (Topbas 2024). Wholesalers contribute to price discovery and execute large orders with minimal price impact, lowering transaction costs for market participants (T. A. Ernst 2026). Credible reporting claims that only three firms dominate 50% of the US retail brokerage industry (T. A. Ernst 2024). As a result, rationalizing the ever-increasing volume and complexity of data has become difficult for a human observer. In September 2026, the site Revenuebase.ai observed 3072 automated trading platforms in the world, with most – 569 in the USA (revenuebase.ai 2026). Surprisingly, India ranks second, ahead of the UK.

The proliferation and growing institutional role of automated market makers (AMMs) specializing in high-frequency trading (HFT), as well as technology-driven liquidity provision, call for regulating these entities and applying traditional regulatory instruments to them (Poser 2021). In this paper, we quantitatively analyze the information that AMMs and broker-dealers publish under Rule 605. The Securities and Exchange Commission (SEC) adopted Rule 605 in 2000 to "help the public compare and evaluate the execution quality among national securities exchanges … market makers and alternative trading systems" (sec.gov/files/34-99679-fact-sheet.pdf 2024). Rule 605 was amended in 2024 to allow more granularity in time-to-execution buckets of the reporting (FINRA

[1] University of Maryland Global Campus, peter.lerner@faculty.umgc.edu, pblerner18@gmail.com

2024). The measurement horizons were extended from below 100 msec. to 5 minutes and longer, while the timestamps in many databases are currently provided on a nanosecond scale.[2]

The information that AMMs must provide monthly includes certain transaction data in a computer-readable data file with standardized fields (see Appendix for description, (APPENDIX B. 2024)). Because of the nature of AMMs, the reporting file typically contains several hundred thousand records, each consisting of 26 standard fields. These fields, among other things, provide information on the wholesaler's price improvement and execution speed.

The paper is structured as follows. It begins with an introduction that discusses the existence and necessity of the Rule 605 datasets. The second section contains a literature review. The third section describes the data in more detail and divides variables into four groups. These were (1) "static" measures of the order volumes – numbers of orders and volumes of different categories of shares, (2) time to execution, (3) spread statistics, and "dynamic" measures of order volumes, for instance, the time at-the-quote. In the third section, we also provide examples of probability distributions for the major factors and argue for their time stability. The stability of statistical distributions suggests using factor correlation matrices as our state variables. The fourth section deals with higher correlations. The fifth section provides an overview of regression analysis. We tried to predict factors in one group (e.g., spreads) from factors in another group (e.g., volumes). Ordinary Least Squares and Long Short-Term Memory (LSTM) neural networks showed negligible predictive power.

In the sixth section, we analyze connections between some factors using planar graphs built from tree analysis of correlations. Graphic analysis allows us to establish stable connections between factors and, ostensibly, the direction of their causal dependence. The seventh section explores the breakpoints in data using Chow tests. Identified breakpoints point to probable changes in trading algorithms either within the firms themselves or in the exchange's trading engine. The eighth section models month-to-month correlation dynamics. The ninth section delves into summary statistics of spread distributions. If we identify effective spreads with the spreads available to Virtu customers, we conclude that the trading platform appropriates most of the gains from price improvement.

[2] This author thinks the nanosecond time scale is the lowest possible because signal propagation in a nanosecond cannot exceed 30 cm, thus limiting the trading server's location to that distance.

## 2. Literature review

Current literature on automated market making (AMM) is vast. We can divide our necessarily incomplete bibliography into three blocks. These are institutional analysis, whole-market analysis, and analysis of trading algorithms and strategies, including order book theories.

On the institutional side, (Brogaard 2018) and (Anand 2021) established a connection between venue choice and execution quality, and (Van Kervel 2025) established a causal link between DMMs and sharply narrower bid-ask spreads. In addition to the already mentioned (Topbas 2024) and (T. A. Ernst 2024), (Brogaard J. 2022), (Bryzgalova 2023), and (Ernst T. 2023), (T. A. Ernst 2026) also contribute.

In the second batch of the literature, we discuss studies on whole markets. Because information on DeFi markets is more current and widely available than for traditional equity, commodity, and bond markets, some authors have proposed studying crypto markets. We note these references with the view that quantitative features of crypto trading can be generalized to traditional financial markets (Malinova 2024, Battalio 2022, Topbas 2024).

Almgren and Chriss developed a quantitative theory of optimal execution. (Almgren 2001). Moreover, Foucault (2005), Bouchaud (2008), and Cont (2010) continued this line of work. Finally, Wang and Obyzhaeva (Obyzhaeva A. A. 2013) introduced an order-book model in their pioneering work. Several papers from the Princeton-University of Maryland collaboration developed an optimal execution approach based on a continuous order-book model, mean-field approximation, and stochastic partial differential equations in detail (see, e.g., (Nutz 2023)).

Among the papers studying empirical bid-ask spreads, which are part of the current paper's analysis, we can mention the following. Kaeck, Van Kervel, and Seeger (Kaeck 2022) evaluated bid-ask spreads for index options. Bessembinder studied distributions of average and effective spreads, and executions within and outside the NBBO (Bessembinder 2020), using NYSE as an example, and (Bellia 2023) did the same for Euronext.

## 3. The data

We used Rule 605 reporting data from the market-making company Virtu Financial, which claims to provide competitive bids and offers in over 25,000 securities across over 200 venues in 37 countries (Virtu.com n.d.). We used its reporting files from 2021 to the beginning of 2024 (39 months, though some data from Jan. 2021 were corrupted in the original file) and analyzed them using the proposed methodology. Each reporting file contains about 200,000 records classified according to the 26 standard categories. Many records contain only zeros or NaN values, so we removed all content from these records. The Appendix provides a detailed list of the categories. From these, we selected four groups indicating depth, execution speed (number of shares in each time interval), spread characteristics, and the proportion of shares whose price improved or stayed at the quote. We deliberately ignored some categories. For example, we did not include stock names because we limited our analysis to company-wide aggregate statistics. Table 1 lists the groups we analyzed.

SoHo Global is a multi-asset online broker based in London that provides a trading platform for institutional and retail clients. Our database reflects the operation of its previous version, SoHo Markets, from January 2021 to the first quarter of 2023.

We were interested in developing a characterization of these data that is 1) relatively stable from month to month and 2) has some predictive power. Our first characterization is the shape of probability distributions for our select quantities.

Tables 2A and 2 B below show summary statistics for the categories for Virtu and SoHo. Figure 1 shows the clear downward trend in dollar volumes over calendar time during the observation years for both companies. We attribute the downward trend in volumes to the steadily increasing number of automated trading platforms worldwide.

The observed tendencies are as follows. Average spreads are volatile and sometimes become negative. Average spreads of Virtu and Soho follow each other quite tightly, except for the “extreme” months, when they diverge significantly. Effective spreads are never negative. SoHo spreads are systematically lower than Virtu spreads. Both companies' spreads demonstrate a weak decaying tendency with calendar time.

## 4. Higher correlations

Higher correlations between trading statistics can be calculated, but several obstacles limit their use in identifying patterns. First, higher correlations can be defined in different non-equivalent ways (correlations between pairs of variables, higher cumulants, etc.). Second, there are too many, making detailed interpretation impossible. Already, for the four groups (see Table 1) of four factors, consisting of correlations between non-repeating pairs of commuting variables, the number is $P^2\binom{6}{4}/4!$=5400. In Table 3, we display non-repeating correlations of correlations according to the following definition:

$$\begin{aligned} \rho_{ij}^{ab}(t) &= Corr[r_i^a r_j^b, r_i^a r_j^b] \\ rho(1)_{ij} &= Corr[\rho_{ij}^{ab}(t_0), \rho_{ij}^{ab}(t_0+12)] \\ rho(2)_{ij} &= Corr[\rho_{ij}^{ab}(t_1), \rho_{ij}^{ab}(t_1+12)] \end{aligned} \quad (1)$$

In Equation (1), $r_{i,j}^{a,b}$ are the factors *a* and *b* from the groups *i* and *j*. There, *i, j*=1→4, *a, b*=1→2, $t_0$ – calendar month in year 0, $t_1$ – calendar month in year 1. Non-repeating correlation combinations for each month (we display the month of April for the years 21/22 and 22/23 in Table3) constitute six combinations (*ij* = 12, 13, 14, 23, 24, 34) because of the commutativity of our factors. Observed fourth-order correlations are highly stable through calendar time ($R^2$≈90-100%), but their significance is hard to interpret.

## 5. Regression analysis

In this section, we attempted to predict some features of one group from the features of another group. The most useful prediction for practitioners is spreads. Table 4, where we used OLS to predict spreads (Group 3) from the static share measures (Group 1), demonstrates that the predictive power of the linear regression is very small. Here, we must caution that the predictive power of the records' regression is not equivalent to the predictive power for the shapes of probability distributions (Figure 1). One can achieve extremely good predictions of the shape of probability distributions using only shifts and affine transformations of the values, with the coefficient of determination sometimes reaching 1. However, using shifts and affine transformations to fit distributions obviates predictability for a practical user.

More sophisticated analysis methods do not change this picture. We used a Long Short -Term Memory neural network (LSTM) to analyze the same dataset. Figure 10 shows schematics of LSTM architecture. The coefficient of determination for LSTM is even smaller than for OLS. Again, the correlation between the shapes of probability distributions can reach unity even without affine transformation, suggesting that the probability distributions are highly predictable. We shall discuss the uses of this fact in subsequent sections.

## 6. Graphic analysis of connections

6.1 Visual analysis of graph correlations

Because correlation matrices are stable, we can analyze the implied connections between selected features of the statistics report. To do this, we used four average correlation matrices for 2021, 2022, fifteen months of 2023-2004, and the entire 39-month observation period. We plotted each matrix as a planar graph. Fig. 10 shows a sample graph built by a spanning tree algorithm for the 2022 averages. Because the algorithm is stochastic, graphs can look different from run to run, but the connectivity persists.

Table 4 presents the results of a visual inspection of persistent connections for Virtu. The conclusions from the graphic analysis are supported by naïve intuition. Cancellations happen mostly in the later periods of execution for a given order. In contrast, most improvements happen in the first seconds and, naturally, are roughly proportional to the number of orders and shares covered.

Tables 5 and 6 provide statistics on persistent correlations between factors identified through visual inspection of graphs. Applying t- and r-statistics (correlation significance) confirms that, in most cases, the correlations implied by the graphical analysis are significant.

The results are as follows. The number of orders is highly correlated with execution time; the first batch (sec09) has the highest positive correlation. Share cancellations mostly occur after the first ten seconds. Canceled shares are anticorrelated with the time batch. Further, order cancellation volume increases over time. The proportion of improved shares is highest in the first ten seconds and lowest thereafter. Shares' spread improvement happens in the first ten seconds and is mostly insignificant thereafter. Shares execute at the quote in the first ten seconds. After the first ten seconds, anticorrelation emerges. Effective spread stabilizes (positive correlation) in the first ten seconds of execution and exhibits a statistically insignificant correlation after that. Price spread improvement happens in the first ten seconds and is barely significant thereafter. Heretofore, we can summarize that most of the action, except for cancellation of shares, happens during the first batch of execution. By contrast, share cancellations happen in the second, third, and fourth batches (from 10 seconds to 15 minutes after the order).

Shares executed outside the quote, as well as shares executed at the quote and price improvements, correlate with covered shares and anticorrelate with canceled shares. Time outside the quote correlates with covered shares and is not correlated with cancellations.

6.2 Subgraph centrality as a measure of determination

A visual analysis of the graphs can be complemented by quantitative parameters computed from them. One such parameter is *subgraph centrality*. Subgraph centrality measures the number of subgraphs connected to a given node. The subgraph centrality *SC(u)* of a node *u* is defined as the weighted sum of closed walks of all lengths originating and terminating at that node (Estrada 2005). The following Equation provides a computable formula:

$$SC(u) = \sum_{k=0}^{\infty} \frac{\mu_k(u)}{k!} \quad (2)$$

Where $\mu_k(u)$ is a spectral measure of the node (vertex). Obviously, if the graph is finite, Equation (2) has finitely many terms. The spectral measure is calculated as a diagonal element of the k-th power of an adjacency matrix *A*:

$$\mu_k(u) = (A^k)_{uu} \quad (3)$$

Different conventions for normalizing subgraph centrality exist, but they do affect a vertex's ranking.

Another measure of factor affinity is the vertex *closeness centrality*. Closeness centrality is the reciprocal of the average path distance to a vertex *u* over all *n-1* reachable nodes. A formal definition is:

$$C(u) = \frac{n-1}{\sum_{v=1}^{n-1} d(u,v)} \quad (4)$$

where d(u,v) is the distance between nodes and n is the number of reachable nodes.

Because the correlation matrices contain numerically small correlations, we truncated them to $|\rho|>0.15$. Amazingly, *SC(u)* and C(u) are very similar for Virtu and SoHo across the three aggregate periods (years 2021, 2022, 2023 and the three months in 2024 for Virtu, and years 2021, 2022 and the first six months of 2023 for SoHo). Table 7 shows the rank ordering of the nodes for Virtu, and Table 8 for SoHo. Based on subgraph centrality and/or the factor's determinative significance, we have the following hierarchy.

For Virtu, the most important factor is average time outside the quote; second is the number of orders; third is the volume of covered shares; and fourth is the volume of executed shares. The factors indicating speed of execution – the batches between zero and 9 seconds, and between ten and 29 seconds – conclude the list with the same subgraph centrality.

For SoHo, the most important factor is the average time at the quote; second is the number of shares covered; and third is usually the number of improved shares. The factors indicating speed of execution – the batches between zero and 9 seconds, and between ten and 29 seconds – conclude the list. Unlike Virtu, where the rank ordering by subgraph centrality and vertex closeness is identical (at least for the dominant vertex), SoHo's rank ordering by these two measures can differ. The connectivity graphs of these two trading firms were the only qualitative difference we observed in their statistics.

## 7. Breakpoint analysis

The analysis in the previous section relied on visually identifying persistent correlations between factors from planar graph representations of correlation matrices. As mentioned in Sections 4 and 5, most correlations remain stable throughout the 38-39 months of measurements. However, some correlations display a visible regime change at the end of the observation period.

To quantify the change, we applied the Chow test to the time series of select monthly correlations. We selected the variables in question following the analysis in Sections 5 and 6. The most stable and significant were correlations of volume and the execution time slot. Henceforth, we have chosen 1) correlations of the number of orders with four (4) time buckets (executions between 0 and 9 seconds, 10 and 29 seconds, 30 and 59 seconds, and between two and fifteen minutes after order arrival) and 2) correlations of volume of canceled shares with the volume of executed shares and the same abovementioned time buckets.

Table 9 displays the correlation statistics for the hypothesized breakpoint in the Virtu time series in August 2023. We display only statistically significant changes. Some correlations moved from statistically significant to statistically insignificant (i.e., presumed zero) after the changeover.

Attributing changes in correlations to trading patterns is difficult because each monthly database includes thousands of stocks. Even if idiosyncratic events affected a substantial number of stocks, they could not influence trading patterns, which remained stable throughout 2021, 2022, and the first half of 2023. We suggest that either Virtu's trading algorithm changed, or market regulations changed in August 2023.

For a shorter SoHo dataset (27-28 months, depending on the series), the same type of analysis predicts a turnover in the trading algorithm on or around January 2022. The results are highly statistically significant (see Table 10). However, the short length limits the validity of the breakpoint tests.

## 8. Dynamics of the correlation matrices

Because correlation matrices between the factors described in Section 3 are quite stable, we can try to predict their dynamics. As a predictor, we estimate the following Equation by ordinary least squares (OLS) through a standard algorithm:

$$\sigma_{t+1} = \sigma_t + \hat{a}_t \, \hat{L} \, \sigma_t + \hat{u}_t g_t + b_t \otimes I + e_{t+1} \tag{5}$$

In the Equation (4), $\sigma_t$ is a 16×16 matrix of the correlations of 16 factors for each month, $\hat{L}$ is a graph Laplacian (Chung 1992), $g_t$ is a 16×16 matrix of graph distances, $\hat{a}_t$ is the estimated 16×16 matrix of volatilities, $\hat{u}_t$ is estimated 16×16 matrix of drift, $b_t$ is the estimated 16×1 vector of regression biases, $I$ is a 16×1 unit vector and $e_{t+1}$ is the regression error used to estimate mean statistical error (MSE).

Despite a tedious form, Equation (4) is simply a discretized linear estimator of a multidimensional diffusion equation – the standard tool of mathematical finance since Black and Scholes (Hull 2017), (Jeanblanc 2009), (Øksendal 2010).[3] In Equation (4), $\hat{a}$ is a diffusion coefficient, and $\hat{u}$ is a monthly factor drift. There is a mathematical argument in favor of replacing the graph Laplacian in (4) with a Ricci curvature. It allows interpretation of the dynamic Equation (4) in terms of surface deformations, but this discussion is beyond the scope of the current paper (Hamilton 1986).

The above regression is highly significant and explains, on average, 28% of the monthly variation in the sixteen factors. However, monthly diffusion and drift coefficients are not statistically persistent from month to month:

$$\frac{1}{38}\sum_{t=1}^{38} \hat{a}_t \approx 0, \frac{1}{38}\sum_{t=1}^{38} \hat{u}_t \approx 0 \tag{5}$$

This approximate equality indicates that the coefficient averages are well within the standard deviation over the 38 months of observation. Table 11 provides detailed statistics supporting Equation (5).

---

[3] Specifically, multidimensional random processes in the context of option pricing were studied by Muisella and Rutkowski (Muisella 1997) and Carmona and Durrelman (Carmona 2003)

## 9. **Analysis of the bid-ask spreads**

Fig. 5A shows the Gaussian kernel-smoothed histogram of the effective bid-ask spread for Virtu. It shows that the fat tail (see below) of the bid-ask spread distribution extends only to the positive side. Compared with Figure 5B, the average bid-ask spread shows an absolute improvement in price.

In Fig. 11, we plot histograms of average spreads for three months of Virtu trading on a logarithmic scale, with tails fitted to power laws with $\alpha_1$=3.0 (positive, right tail) and $\alpha_2$=2.75 (negative, left tail).[4] On a logarithmic scale, they are almost symmetric. The explanatory variance of the smoothed average spreads of Virtu distributions for the average spread distributions of SOHO for the same three months exceeds 99%.

A positive tail of the effective spread is very close to the positive tail of the average spread. In contrast, the negative tail of the effective spread is negligible. If we identify the effective spread with the spread of the trading system's customers, we summarize the above observations as "in the game with the customer/broker, AMMs (almost) never lose".

## **Conclusion**

Our analysis of the Rule 605 reports of Virtu and SoHo during 2021-2024 showed the following stylized facts. The distributions of the 16 reported factors are highly stable over time and highly correlated between the two trading firms. Next month's factor correlations can be accurately predicted ($R^2$~98%÷99%) by using the current month's data. This is not proof, but it strongly suggests that drift in parameter correlations outside of breakpoints is a martingale.

About a quarter of the monthly change (first difference) in sixteen factors can be explained by past volatility and factor drift. The variability of average spreads is higher than that of effective spreads. The distribution of average spreads has a negative tail indicating absolute price improvement, not just improvement within the bid-ask spread. The distribution of effective spreads has only a positive tail and is truncated at zero, suggesting that market makers almost always trade at a profit. Note that if the trading strategies were not heavily skewed in favor of AMMs, the volume of trades in favor of the AMM and the client must have been comparable. This requires further investigation because, given the very short time between placing an order and its execution, one cannot imagine adverse selection (informed trading) occurring in milliseconds.

[4] Unexpectedly, the author observed the same tail exponent of 3 in the bid-ask spread for an entirely different trading book (Lerner 2009).

Most price improvement happens within the first 10 seconds. Most cancellations happen substantially later. This is a natural result – most matching orders tend to be executed as fast as possible to prevent slippage – and most hanging orders stay until being canceled.

Connectivity studies mostly agree with this picture. For Virtu, the most connected node is the average time outside the quote; the number of orders and volume measures follow it in terms of connectivity. For SoHo, the most connected node is average time at the quote, and the volume metrics are similar to Virtu's.

Our correlation analysis identified a structural break in Virtu data around August 2023 and in SoHo data around January 2022. Because thousands of assets trade on each platform, we cannot attribute the structural break to any single market event. It could be explained by a changeover in the stock market or a changeover of trading algorithms within these companies. Because these breaks occurred at different times for Virtu and SoHo, we cautiously propose a second explanation. However, it would be interesting to interview members of the trading firms about these observations.

# Appendix

U.S. Securities and Exchange Commission

## APPENDIX B

**Table with Rule 11Ac1-5 Sample Statistics for a Single Security "A" in May 2001**

| F1 | F2 | F3 | F4 | F5 | F6 | F7 | F8 | F9 | F10 | F11 | F12 | F13 | F14 | F15 | F16 | F17 | F18 | F19 | F20 | F21 | F22 | F23 |
|---|---|---|---|---|---|---|---|---|---|---|---|---|---|---|---|---|---|---|---|---|---|---|
| T | MM1 | 200105 | A | 11 | 21 | 106 | 22200 | 400 | 21500 | 300 | 19200 | 1300 | 800 | 300 | 200 | .0463 | .0852 | 1900 | .0389 | 12.4 | 18800 | 6.3 |
| T | MM1 | 200105 | A | 11 | 22 | | | | | | | | | | | | | | | | | |
| T | MM1 | 200105 | A | 11 | 23 | | | | | | | | | | | | | | | | | |
| T | MM1 | 200105 | A | 11 | 24 | | | | | | | | | | | | | | | | | |
| T | MM1 | 200105 | A | 12 | 21 | | | | | | | | | | | | | | | | | |
| T | MM1 | 200105 | A | 12 | 22 | | | | | | | | | | | | | | | | | |
| T | MM1 | 200105 | A | 12 | 23 | | | | | | | | | | | | | | | | | |
| T | MM1 | 200105 | A | 12 | 24 | | | | | | | | | | | | | | | | | |
| T | MM1 | 200105 | A | 13 | 21 | | | | | | | | | | | | | | | | | |
| T | MM1 | 200105 | A | 13 | 22 | | | | | | | | | | | | | | | | | |
| T | MM1 | 200105 | A | 13 | 23 | | | | | | | | | | | | | | | | | |
| T | MM1 | 200105 | A | 13 | 24 | | | | | | | | | | | | | | | | | |
| T | MM1 | 200105 | A | 14 | 21 | | | | | | | | | | | | | | | | | |
| T | MM1 | 200105 | A | 14 | 22 | | | | | | | | | | | | | | | | | |
| T | MM1 | 200105 | A | 14 | 23 | | | | | | | | | | | | | | | | | |
| T | MM1 | 200105 | A | 14 | 24 | | | | | | | | | | | | | | | | | |
| T | MM1 | 200105 | A | 15 | 21 | | | | | | | | | | | | | | | | | |
| T | MM1 | 200105 | A | 15 | 22 | | | | | | | | | | | | | | | | | |
| T | MM1 | 200105 | A | 15 | 23 | | | | | | | | | | | | | | | | | |
| T | MM1 | 200105 | A | 15 | 24 | | | | | | | | | | | | | | | | | |

F1: Designated Participant Code
F2: Market Center Code
F3: Date
F4: Security Ticker Symbol
F5: Order Type Code
F6: Order Size Code
F7: Total Covered Orders
F8: Total Covered Shares
F9: Cancelled Shares

F10: Market Center Executed Shares
F11: Away Executed Shares
F12: Shares from 0 to 9 Seconds
F13: Shares from 10 to 29 Seconds
F14: Shares from 30 to 59 Seconds
F15: Shares from 60 to 299 Seconds
F16: Shares from 5 to 30 Minutes
F17: Average Realized Spread ($)
F18: Average Effective Spread ($)

F19: Price-Improved Shares
F20: Price Improved Average Amount ($)
F21: Price Improved Average Time (seconds)
F22: At-the Quote Shares
F23: At-the-Quote Average Time (seconds)
F24: Outside-the-Quote Shares
F25: Outside-the-Quote Average Amount ($)
F26: Outside-the-Quote Average Time (seconds)

*http://www.sec.gov/interps/legal/slbim12b.htm*

Table 1. Factors of the report used in the analysis of Rule 605 reporting. Sixteen factors are numbered from 0 to 15 according to the Python convention.

| | |
|---|---|
| **Group 1: Static measures of stock quantities.** | |
| **a.** | Number of orders (‘covord’) |
| **b.** | Number of covered shares (‘covsh’) |
| **c.** | Number of canceled shares (‘cansh’) |
| **d.** | Number of executed shares (‘exsh’) |
| **Group 2: Execution time measures.** | |
| **a.** | Shares executed in less than 10 seconds ('sec09') |
| **b.** | Shares executed between 10 and 29 seconds ('sec1029') |
| **c.** | Shares executed between 30 and 59 seconds (‘sec3059’) |
| **d.** | Shares executed between 2 minutes and 15 minutes (‘min215’) |
| **Group 3: Measures of spread.** | |
| **a.** | Average spread (‘avrsp’, $) |
| **b.** | Effective spread (‘effsp’, $) |
| **c.** | Shares spread improvement (‘impr’) |
| **d.** | Price spread improvement (‘primp’, $) |
| **Group 4: Dynamic measures of stock quantities.** | |
| **a.** | Shares executed outside of quote (‘outq’) |
| **b.** | Shares executed at the quote (‘atq’) |
| **c.** | Shares executed with improvement in price (‘imsh’) |
| **d.** | Average seconds outside the quote (‘secout’, sec.) |

Table 2A. Summary statistics of trading for Virtu for the years 2021-2023 and the first three months of 2024.

| | 2021 | 2022 | 2023 | 23/24 (16 mo.) |
|---|---|---|---|---|
| Avg # records | 195570 | 203966 | 172270 | 1.71E+05 |
| **Group 1** | | | | |
| covsh | 3.49E+10 | 3.16E+10 | 2.28E+10 | 2.24E+10 |
| exsh | 1.80E+10 | 1.89E+10 | 1.13E+10 | 1.10E+10 |
| cansh | 1.69E+10 | 1.27E+10 | 1.15E+10 | 1.14E+10 |
| **Group 2** | | | | |
| sec09 | 1.61E+10 | 1.18E+10 | 1.08E+10 | 1.07E+10 |
| sec1029 | 1.93E+08 | 1.73E+08 | 1.53E+08 | 1.53E+08 |
| sec3059 | 1.23E+08 | 1.04E+08 | 9.54E+07 | 9.45E+07 |
| min215 | 2.55E+08 | 2.05E+08 | 1.97E+08 | 1.94E+08 |
| **Group 3** | | | | |
| avsp | 1.58E-02 | 1.56E-02 | 1.46E-02 | 1.48E-02 |
| effsp | 3.23E-02 | 2.60E-02 | 2.67E-02 | 2.69E-02 |
| impr | 9.06E-03 | 1.30E+02 | 8.97E-03 | 9.26E-03 |
| primp | 6.73E-03 | 4.71E-03 | 4.64E-03 | 4.72E-03 |
| **Group4** | | | | |
| secat | 2.26E+03 | 1.51E+03 | 1.69E+03 | 1.73E+03 |
| outquote | 4.10E+09 | 2.32E+09 | 2.07E+09 | 2.09E+09 |
| amotq | 1.12E+10 | 8.94E+09 | 8.11E+09 | 7.99E+09 |
| secout | 6.85E+00 | 6.22E+00 | 6.18E+00 | 6.08E+00 |

Table 2B. Summary statistics of trading by SoHo in 2021-2023.

| | 2021 | 2022 | 2023 |
|---|---|---|---|
| Avg # records | 124636 | 122219 | 115389 |
| **Group 1** | | | |
| covsh | 8.2E+09 | 6.16E+09 | 5.22E+09 |
| exsh | 4.06E+09 | 3.58E+09 | 2.89E+09 |
| cansh | 3.98E+09 | 2.46E+09 | 2.23E+09 |
| **Group 2** | | | |
| sec09 | 3.17E+09 | 1.88E+09 | 1.77E+09 |
| sec1029 | 2.3E+08 | 1.61E+08 | 1.06E+08 |
| sec3059 | 1.43E+08 | 96458227 | 66507365 |
| min215 | 2.74E+08 | 1.81E+08 | 1.32E+08 |
| **Group 3** | | | |
| avsp | 0.017855 | 0.015894 | 0.009589 |
| effsp | 0.025082 | 0.020071 | 0.019255 |
| impr | 9.85E-03 | 7.99E-03 | 7.85E-03 |
| primp | 4.56E-03 | 2.90E-03 | 2.51E-03 |
| **Group4** | | | |
| secat | 1509.583 | 733.7353 | 640.4665 |
| outquote | 9.98E+08 | 5.01E+08 | 3.96E+08 |
| amotq | 1.31E+09 | 8.85E+08 | 1.05E+09 |
| secout | 3.578333 | 2.647652 | 2.087327 |

Table 3. Coefficients of determination $R_{ij}^2=(rho(t)_{IJ})^2$ for the fourth-degree correlations of factors between April in years 2021, 2022 and 2023 for Virtu.

| | YEARS | |
|---|---|---|
| ***rho(*t)$_{IJ}$** | 2021/22 | 2022/23 |
| 12 | 0.9994 | 0.9993 |
| 13 | 0.9407 | 0.9037 |
| 14 | 0.9960 | 0.9973 |
| 23 | 0.9622 | 0.9873 |
| 24 | 0.9983 | 0.9997 |
| 34 | 0.9591 | 0.8963 |

Table 4. Adjusted explained variance for the linear regression of spreads (Group 3) from the preceding trading volumes (Group 1) for Virtu. For brevity, we provide only data for the first 15 months in the sample.

| **Records** | **Month** | **Linear Regression Explained Variance** | **LSTM Explained Variance** | **Notes** |
|---|---|---|---|---|
| **1** | January | - | - | Partially corrupted Virtu dataset |
| **2** | February | - | - | Not predictable from January's data |
| **3** | March | 0.4053% | 0.0000% | |
| **4** | April | 0.3650% | -0.0142% | |
| **5** | May | 0.5333% | 0.0003% | |
| **6** | June | 0.3499% | 0.0001% | |
| **7** | July | 0.4049% | 0.0000% | |
| **8** | August | 0.4837% | 0.0000% | |
| **9** | September | 0.3445% | -0.0244% | |
| **10** | October | 0.7374% | 0.0035% | |
| **11** | November | -0.1954% | -0.0007% | |
| **12** | December | 0.3687% | -0.0002% | |
| **13** | January | 1.0505% | 0.0009% | |
| **14** | February | -0.1236% | -0.0796% | |
| **15** | March | 1.1044% | 0.0000% | |

Table 5. Persistent connectivity of Virtu factor correlation matrices in the average samples of 2021, 2022, and 2023-2024 (16 months), averaged over the three years.

| **Central determinant: Group** | **Cluster: Group** | **Notes** |
|---|---|---|
| Canceled shares(cansh): 1 | sec10-29: **2**, sec30-59: **2**, secout: **4** | Cancellation of shares happens mostly beyond 10 seconds after an order submission/ |
| Price improvement (primp): 3 | covord: **1**<br>sec09: **2** | Price improvement happens within the first 10 seconds. Price improvement correlates with the number of orders. |

Table 6A. Persistent correlations of the Virtu factors from groups 1, 3 and 4 with group 2 (speed of execution) factors. Standard deviation is shown in parentheses. See Table 5B for the correlation statistic.

**Average (std)**

| Correlation with (group) factor | sec09 | sec1029 | sec3059 | min25 |
|---|---|---|---|---|
| (1) covord | 0.5736<br>(0.0626) | 0.3150<br>(0.0263) | 0.3102<br>(0.0226) | 0.2918<br>(0.0253) |
| (1) cansh | -0.1113<br>(0.0385) | 0.2583<br>(0.0270) | 0.2781<br>(0.0259) | 0.3010<br>(0.0304) |
| (4) imsh | 0.6178<br>(0.0219) | -0.1074<br>(0.0251) | -0.1598<br>(0.0186) | -0.2064<br>(0.0262) |
| (3) impr | 0.1192<br>(0.0261) | -0.0039<br>(0.0562) | -0.0428<br>(0.0124) | -0.0596<br>(0.0141) |
| (4) atq | 0.6723<br>(0.0200) | -0.0855<br>(0.0222) | -0.1378<br>(0.0180) | -0.1778<br>(0.0233) |
| (3) effsp | 0.0873<br>(0.0313) | -0.0160<br>(0.0097) | -0.0330<br>(0.0095) | -0.0330<br>(0.0095) |
| (3) primp | 0.1018<br>(0.0357) | 0.0127<br>(0.0078) | -0.0146<br>(0.0063) | -0.0266<br>(0.0089) |

Table 6B. Virtu statistics of correlations of the factors from groups 1, 3, and 4 with group 2 (speed of execution). Each cell shows the t- and r-statistics, separated by "/".Most correlations are highly statistically significant (P<0.1%) and indicate that most actions, except share cancellation, occur in the first 10 seconds of execution.

| **Student t/r-test** | | | | | **Comments** |
|---|---|---|---|---|---|
| (g**roup**) factor | sec09 | sec1029 | sec3059 | min25 | |
| **(1) covord** | 9.1696/ 25.9088 | 11.9911/ 12.2788 | 13.7561/ 12.0737 | 11.5225/ 11.2860 | The number of orders is highly correlated with execution time; the first batch (sec09) has the highest correlation. |
| **(1) cansh** | -2.8918 -4.1429 | 9.5830 9.8941 | 10.7462 10.7126 | 9.9117 11.6766 | Share cancellations mostly happen after the first 10 seconds (anticorrelation). Cancellations increase with time. |
| **(4) imsh** | 28.2406/ 29.0667 | -4.2854/ -3.9988 | -8.5991/ -5.9892 | -7.8715/ -7.8066 | The proportion of improved shares is the highest in the first 10 seconds and the lowest (anticorrelation) after that. |
| **(3) impr** | 4.5697/ 4.4438 | -0.0701/ -0.1458 | -3.4502/ -1.5852 | -4.2215/ -2.2101 | Share spread improves in the first 10 seconds and is mostly insignificant thereafter. |
| **(4) atq** | 33.5934/ 33.5983 | -3.8514/ -3.1753 | -7.6357/ -5.1483 | -7.6480/ -6.6861 | Shares are executed at the quote in the first 10 seconds. After 10 seconds, anticorrelation occurs. |
| **(3) effsp** | 2.7857/ 3.2424 | -1.6441/ -0.5912 | -3.4625/ -1.2224 | -2.2875/ -1.6408 | Effective spread stabilizes in the first 10 seconds of execution. Statistically insignificant correlation after 10 seconds. |
| **(3) primp** | 2.8563/ 3.7877 | 1.6371/ 0.4699 | -2.3113/ -0.5409 | -2.9796/ -0.9846 | Price spread improvement happens in the first 10 seconds and is barely significant thereafter. |

Table 7A. Persistent correlations of the factors from groups 3 and 4 with group 1 (logarithms of trading volume). Most of the correlations are highly statistically significant (P<0.1%). Standard deviation is shown in parentheses.

**Factor average (std)**

| **Correlation with (group) factor** | covsh | cansh | exsh |
|---|---|---|---|
| **(4) outq** | 0.3962 (0.0204) | -0.2049 (0.0401) | 0.4509 (0.0160) |
| **(4) atq** | 0.3664 (0.0310) | -0.1849 (0.0415) | 0.5623 (0.0230) |
| **(3) primp** | 0.2121 (0.0435) | -0.2913 (0.0261) | 0.5020 (0.0262) |
| **(4) secout** | 0.1245 (0.0225) | 0.0002 (0.0189) | 0.1507 (0.0142) |

Table 7B. Persistent correlations of the factors from groups 3 and 4 with group 1 (logarithms of trading volumes). Results of the Student's t-test and correlation significance test are shown, separated by "/".

| **Student t/r-test** | | | | **Comments** |
|---|---|---|---|---|
| | covsh | cansh | exsh | |
| **(4) outq** | 19.3928/ 15.9660 | -5.1098/ -7.7449 | 28.2339/ 18.6894 | Shares executed outside of quote are correlated with covered and anticorrelated with canceled shares. |
| **(4) atq** | 11.8226/ 14.5681 | -4.4592/ -6.9595 | 24.4046/ 25.1620 | Shares executed at the quote correlate with covered shares and anticorrelate with canceled shares. |
| **(3) primp** | 4.8750/ 8.0301 | -11.1463/ -11.2674 | 19.1242/ 21.4743 | Price-improved shares correlated with the covered shares and anticorrelated with the canceled shares. |
| **(4) secout** | 5.5207/ 4.6414 | 0.0089/ 0.0062 | 10.5973/ 5.6421 | Time outside the quote correlated with covered shares but not with cancellations. |

Table 8. Rank ordering of the importance of the dominant Virtu correlations for their subgraph centrality SC(u) and distance closeness C(u). They are the same for each of the three combined periods in the database (2021-A), (2022-B), and (2023-three months of 2024-C). We observe that they are mostly consistent between consecutive years.

A)

| **Virtu factors, 2021** | **SC(u)** | **C(u)** |
|---|---|---|
| secout (15) | 0.137552 | 1 |
| covord (0) | 0.111896 | 0.789474 |
| covsh (1) | 0.111896 | 0.789474 |
| exsh (3) | 0.099095 | 0.75 |
| sec09 (4) | 0.085268 | 0.714286 |
| imsh (14) | 0.06914 | 0.714286 |

B)

| **Virtu factors, 2022** | **SC(u)** | **C(u)** |
|---|---|---|
| secout (15) | 0.13629 | 1 |
| covord (0) | 0.115591 | 0.789474 |
| covsh (1) | 0.115591 | 0.789474 |
| sec09 (4) | 0.102135 | 0.75 |
| imsh (14) | 0.087621 | 0.714286 |

C)

| Virtu factors, 2023/4 | SC(u) | C(u) |
|---|---|---|
| secout (15) | 15 | 0.145684 |
| covord (0) | 1 | 0.122288 |
| covsh (1) | 3 | 0.107477 |
| exsh (3) | 0 | 0.10667 |
| sec09 (4) | 4 | 0.076906 |
| sec1029 (5) | 5 | 0.075365 |

Table 9. Rank ordering of the importance of the dominant SoHo correlations for their subgraph centrality SC(u) and distance closeness C(u). They are the same for each of the three combined periods in the database (2021, 2022, 2023 (6 mo.)). Unlike the Virtu correlations, the rank ordering by subgraph centrality and graph distance slightly differs between factors.

| SOHO (2021). | | | SOHO, 2022 | | | SOHO, 2023 (6 mo.) | | |
|---|---|---|---|---|---|---|---|---|
| | SC(u) | C(u) | | SC(u) | C(u) | | SC(u) | C(u) |
| atq (13) | 0.115842 | 0.75 | atq (13) | 0.116331 | 0.75 | atq (13) | 0.112178 | 0.625 |
| covsh (1) | 0.10811 | 0.714286 | covsh (1) | 0.111282 | 0.652174 | covsh (1) | 0.112178 | 0.625 |
| imsh (14) | 0.105454 | 0.833333 | imsh (14) | 0.096241 | 0.714286 | sec09 (4) | 0.097795 | 0.5 |
| sec3059 (6) | 0.091965 | 0.652174 | sec1029 (5) | 0.094856 | 0.6 | sec1029 (5) | 0.097795 | 0.5 |
| sec1029 (5) | 0.091965 | 0.652174 | sec3059 (6) | 0.094856 | 0.6 | sec3059 (6) | 0.093892 | 0.6 |
| | | | | | | imsh (14) | 0.093892 | 0.6 |

Table 10. Chow breakpoint statistics for Virtu for August 2023 ($32^{nd}$ month into 39-month sample). A star indicates a coefficient of determination for the Chow statistic above 15%. The correlation breakdown for 08/2023 is most prominent for execution in the first 10 seconds after placing an order.

| **Correlations** | | **covord-execution time** | | | | **cansh-exsh** | **cansh-execution time** | | | |
|---|---|---|---|---|---|---|---|---|---|---|
| **break_month** | 32 | **sec09** | **sec1029** | **sec3059** | **min215** | **sec09** | **sec09** | **sec1029** | **sec3059** | **min215** |
| $R^2$ | | 0.5471* | 0.0336 | 0.0400 | 0.1529* | 0.5288* | 0.3218* | 0.0343 | 0.0098 | 0.0373 |
| Mean average | | 0.0302 | 0.0207 | 0.0187 | 0.0184 | 0.0185 | 0.0236 | 0.0204 | 0.0198 | 0.0229 |
| Mean squared error | | 0.0017 | 0.0007 | 0.0005 | 0.0005 | 0.0006 | 0.0009 | 0.0007 | 0.0007 | 0.0009 |
| Median absolute error | | 0.0206 | 0.0166 | 0.0172 | 0.0147 | 0.0165 | 0.0173 | 0.0196 | 0.0178 | 0.0189 |

Table 11. Chow breakpoint statistics for SoHo for January 2022 (11th month into a 28-month sample). A star indicates a Chow statistic coefficient of determination above 15%. The correlation breakdown for 01/2022 is most prominent for execution in the first 10 seconds after placing an order.

| **Correlations** | | **covord-execution time** | | | | **cansh-exsh** | **cansh-execution time** | | | |
|---|---|---|---|---|---|---|---|---|---|---|
| **break_month** | 11 | **sec09** | **sec1029** | **sec3059** | **min215** | **sec09** | **sec09** | **sec1029** | **sec3059** | **min215** |
| $R^2$ | | 0.633* | 0.748* | 0.759* | 0.723* | 0.588* | 0.527* | 0.745* | 0.741* | 0.754* |
| Mean average | | 0.0132 | 0.0146 | 0.0139 | 0.0129 | 0.0161 | 0.0135 | 0.0132 | 0.0127 | 0.0128 |
| Mean squared error | | 0.00037 | 0.00032 | 0.00027 | 0.00023 | 0.00045 | 0.00031 | 0.00034 | 0.00034 | 0.00034 |
| Median absolute error | | 0.00739 | 0.01248 | 0.01301 | 0.01147 | 0.01333 | 0.00988 | 0.00814 | 0.01048 | 0.01077 |

Table 12. Virtu t-Statistics of the matrix of sigma coefficients from Equations (4) and (5), $t = \frac{E[\#]}{Var\ [\#]}$, where # is a placeholder for the regression coefficient between factors $i$ and $j$. None of the coefficients were statistically significant. From this table, we can assume that correlations between factors outside the breakpoints behave as martingales.

| Factors | covord | covsh | cansh | exsh | sec09 | sec1029 | sec3059 | min15 | avrsp | effsp | impr | primp | outq | atq | imsh | secout |
|---|---|---|---|---|---|---|---|---|---|---|---|---|---|---|---|---|
| covord | 0.0596 | 0.0235 | -0.1680 | -0.0971 | 0.0581 | -0.1190 | -0.0474 | 0.0186 | -0.0006 | 0.0699 | -0.0512 | 0.0003 | -0.1920 | -0.2522 | 0.0064 | -0.1029 |
| covsh | -0.2027 | 0.3422 | -0.1155 | 0.1746 | 0.1766 | -0.0417 | 0.0176 | -0.0570 | 0.0491 | -0.1629 | -0.0372 | 0.0018 | 0.1975 | -0.0880 | 0.2554 | -0.1404 |
| cansh | 0.3389 | -0.1037 | -0.1413 | -0.2066 | -0.0893 | -0.3414 | -0.3352 | -0.1499 | -0.0425 | 0.0548 | -0.0560 | -0.0524 | -0.0773 | -0.0962 | -0.1340 | -0.0082 |
| exsh | 0.2589 | -0.0968 | -0.0929 | -0.1911 | 0.0341 | -0.1427 | -0.1737 | -0.0913 | -0.0373 | 0.0418 | -0.0247 | 0.1060 | -0.2101 | 0.0222 | -0.0424 | -0.0546 |
| sec09 | -0.0767 | -0.1690 | 0.0714 | 0.1498 | 0.0806 | 0.0295 | 0.0148 | 0.0160 | 0.1684 | -0.1407 | 0.2463 | 0.3279 | -0.1316 | 0.1581 | 0.0188 | -0.1279 |
| sec1029 | -0.0981 | -0.0836 | 0.1195 | 0.0647 | 0.1117 | 0.1891 | -0.0757 | -0.1950 | 0.0276 | -0.3957 | 0.2631 | 0.0708 | -0.1065 | 0.0401 | 0.0400 | 0.1050 |
| sec3059 | -0.0712 | -0.1211 | 0.1130 | 0.0779 | -0.0091 | 0.2300 | -0.0752 | -0.1360 | 0.0618 | -0.3098 | 0.1874 | -0.1076 | 0.0205 | 0.0284 | 0.0363 | 0.0564 |
| min15 | 0.1526 | 0.1659 | -0.1271 | -0.0279 | 0.0424 | -0.0156 | 0.0087 | -0.1387 | -0.4198 | -0.1082 | 0.1153 | 0.0947 | 0.2311 | -0.2391 | -0.1558 | -0.2164 |
| avrsp | -0.1092 | 0.2415 | 0.1699 | -0.2388 | -0.4994 | 0.1225 | 0.0154 | 0.4666 | 0.2551 | -0.2536 | 0.0101 | -0.3521 | 0.3671 | -0.1730 | 0.4626 | 0.1487 |
| effsp | -0.1702 | -0.1338 | 0.0445 | -0.1031 | -0.3450 | -0.2590 | -0.0264 | 0.0172 | 0.4034 | 0.0341 | -0.1687 | -0.1779 | -0.0582 | 0.1718 | 0.2367 | 0.3193 |
| impr | -0.2314 | 0.0753 | 0.1374 | -0.0034 | -0.1244 | 0.2407 | 0.1157 | -0.0159 | 0.3959 | -0.0372 | 0.1065 | -0.1351 | -0.2175 | -0.0252 | 0.4378 | 0.0150 |
| primp | 0.2454 | -0.2520 | 0.1780 | -0.1371 | -0.0105 | -0.1550 | -0.3183 | -0.1727 | 0.2097 | 0.1318 | 0.1828 | -0.0898 | -0.3371 | -0.1540 | -0.3507 | -0.1728 |
| outq | 0.2903 | -0.1040 | -0.0250 | -0.1848 | 0.0813 | 0.0471 | -0.0580 | 0.0726 | 0.0348 | 0.0311 | -0.1059 | 0.1263 | -0.1892 | -0.0336 | 0.0360 | -0.1217 |
| atq | 0.2763 | -0.0812 | 0.0079 | -0.1328 | 0.0631 | 0.0569 | -0.0864 | 0.0520 | -0.0620 | 0.1476 | -0.1004 | 0.1351 | -0.2187 | 0.1055 | -0.0337 | -0.0541 |
| imsh | 0.1222 | -0.2583 | -0.0010 | 0.0164 | -0.0300 | 0.1955 | -0.0554 | 0.1647 | 0.1068 | 0.1320 | 0.0957 | 0.1745 | -0.2791 | 0.1017 | -0.0987 | -0.0624 |
| secout | -0.0501 | 0.0579 | -0.0089 | -0.1566 | 0.0645 | 0.0327 | -0.0725 | -0.0635 | 0.2076 | -0.2697 | -0.2257 | 0.0928 | -0.1939 | 0.0078 | -0.1319 | -0.3258 |

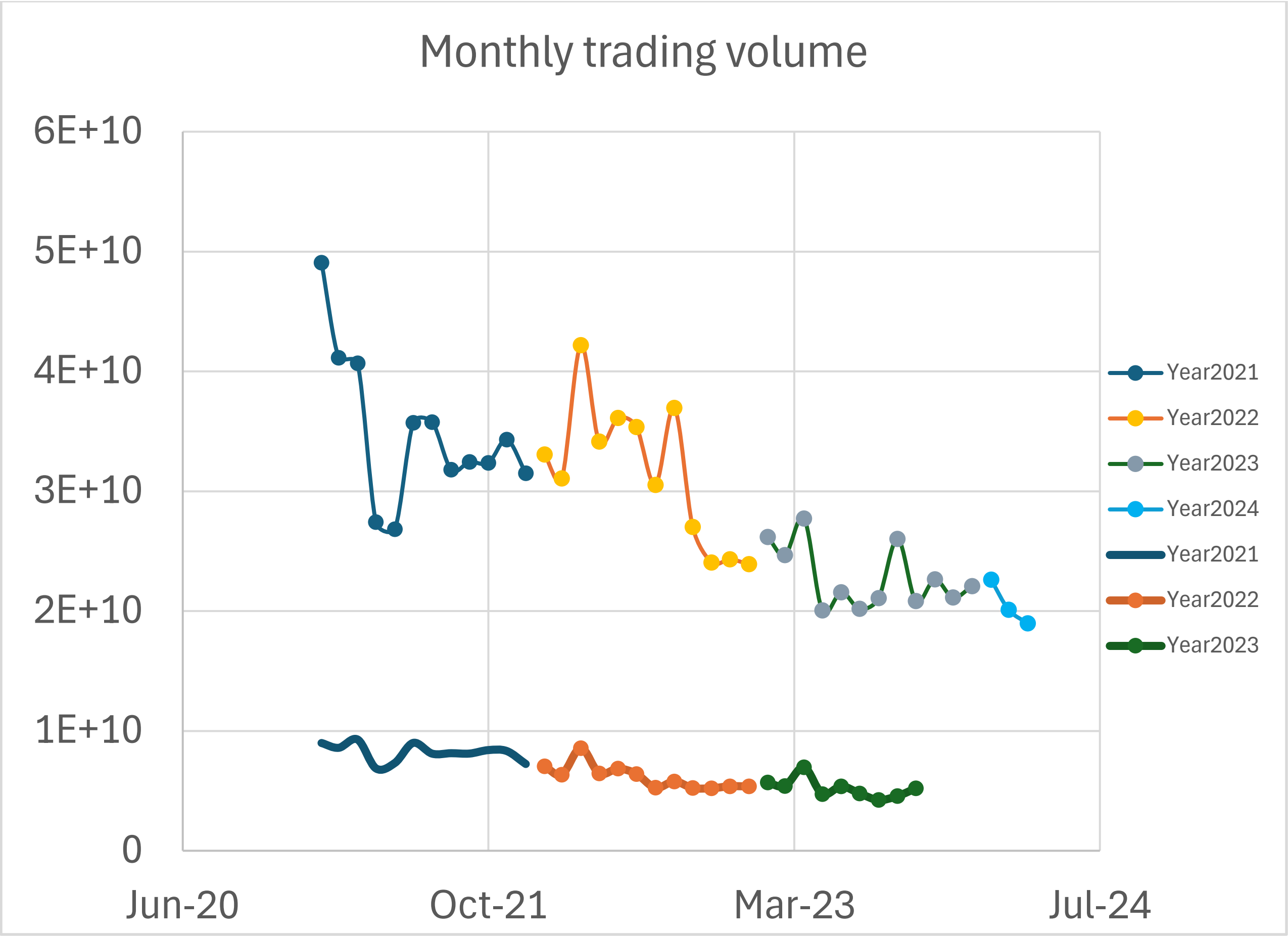

Fig.1 Monthly trading volume in USD. Upper thin curves are for Virtu volumes; lower thick curves are for SoHo volumes. The year-to-year fall in trading volume is almost monotonic.

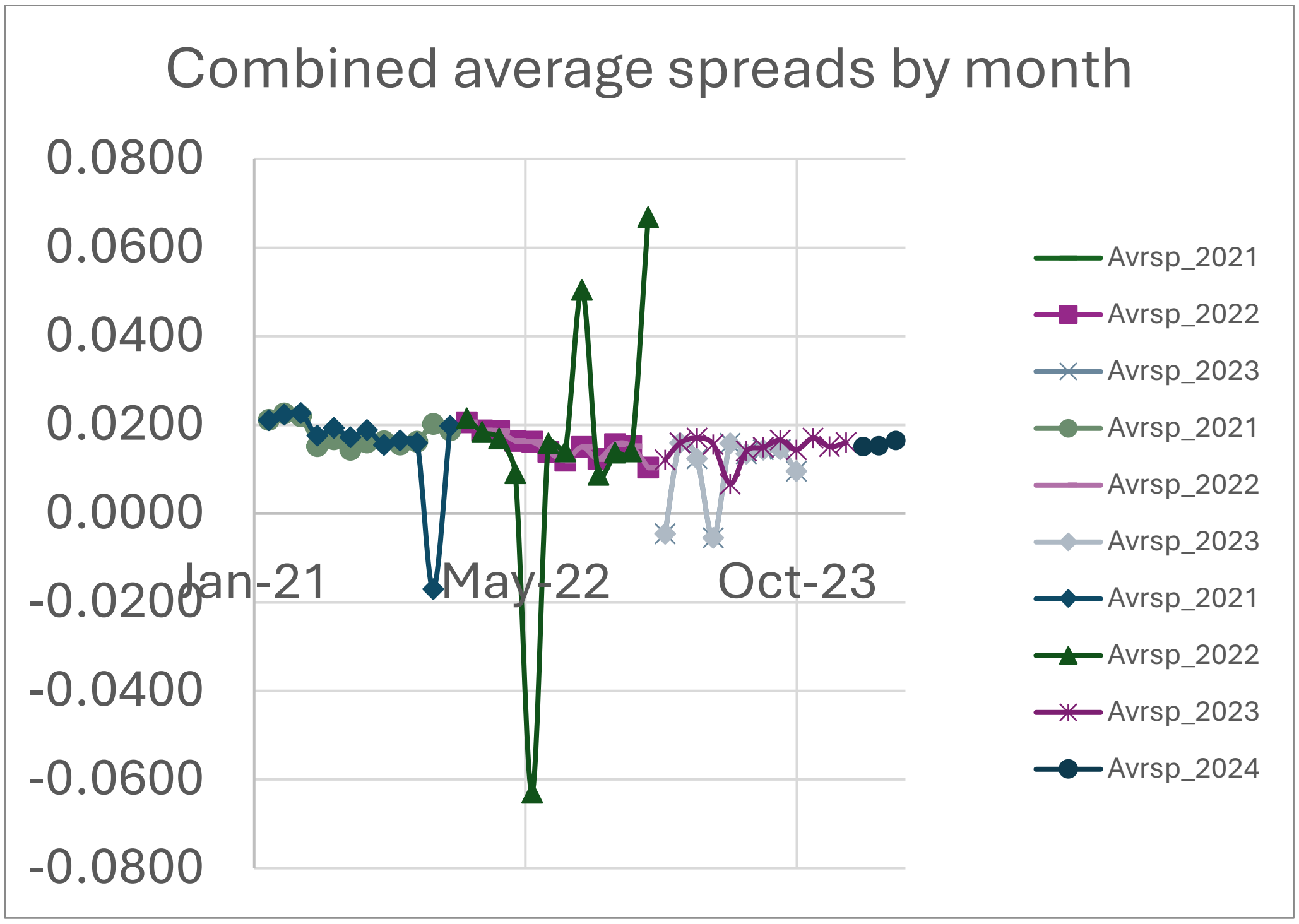
Combined average spreads by month
0.0800
0.0600
0.0400
0.0200
0.0000
-0.0200
-0.0400
-0.0600
-0.0800
Jan-21
May-22
Oct-23
Avrsp_2021
Avrsp_2022
Avrsp_2023
Avrsp_2021
Avrsp_2022
Avrsp_2023
Avrsp_2021
Avrsp_2022
Avrsp_2023
Avrsp_2024

A)

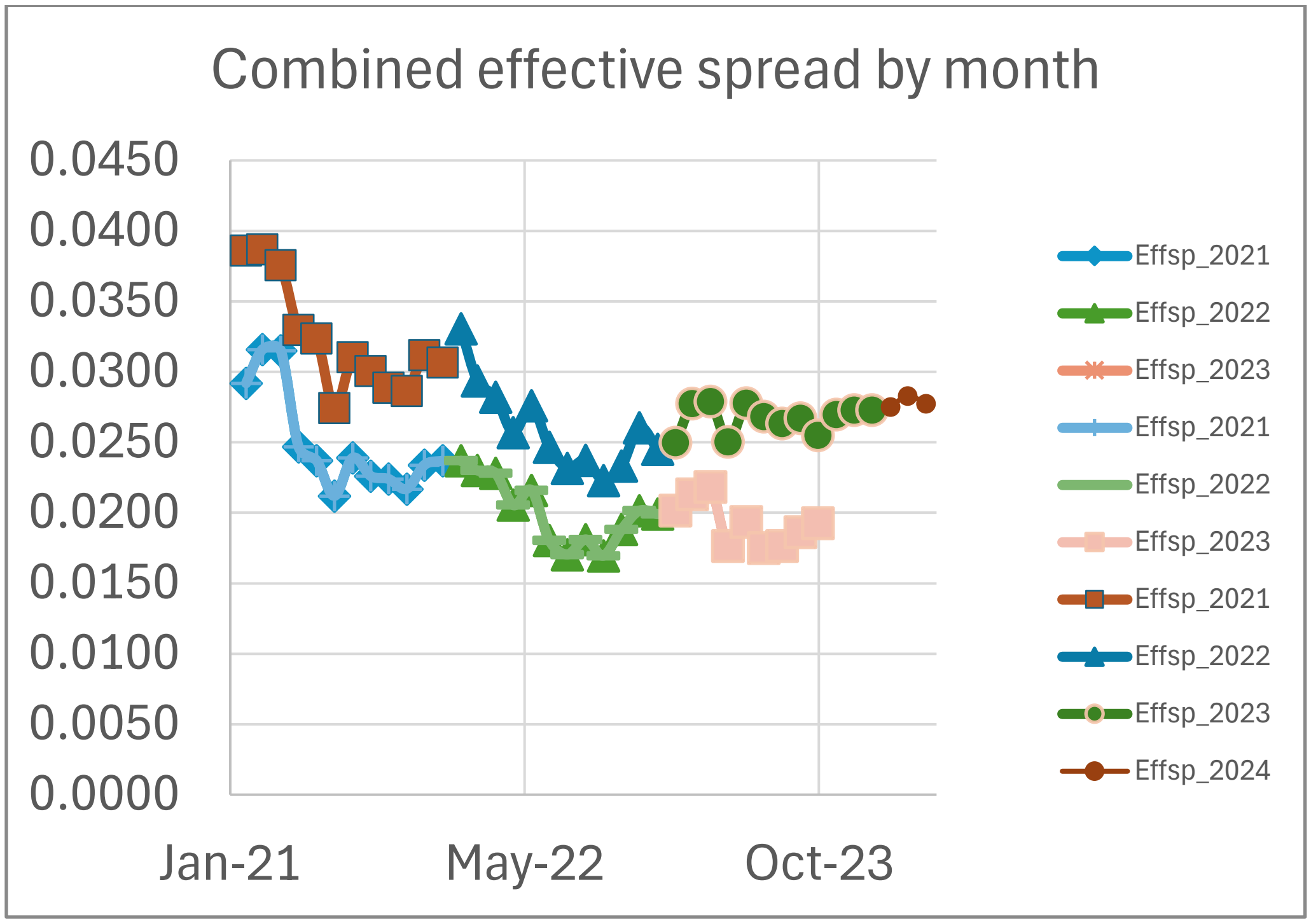
Combined effective spread by month
0.0450
0.0400
0.0350
0.0300
0.0250
0.0200
0.0150
0.0100
0.0050
0.0000
Jan-21
May-22
Oct-23
Effsp_2021
Effsp_2022
Effsp_2023
Effsp_2021
Effsp_2022
Effsp_2023
Effsp_2021
Effsp_2022
Effsp_2023
Effsp_2024

B)

Fig. 2 A) Average trading spreads by month in USD. Average spreads are volatile and sometimes become negative. Virtu and Soho spreads track each other closely, except in the "extreme" months. B) Effective spreads are never negative. SoHo spreads are systematically lower than Virtu's. Both companies' spreads demonstrate a weak decaying tendency with calendar time.

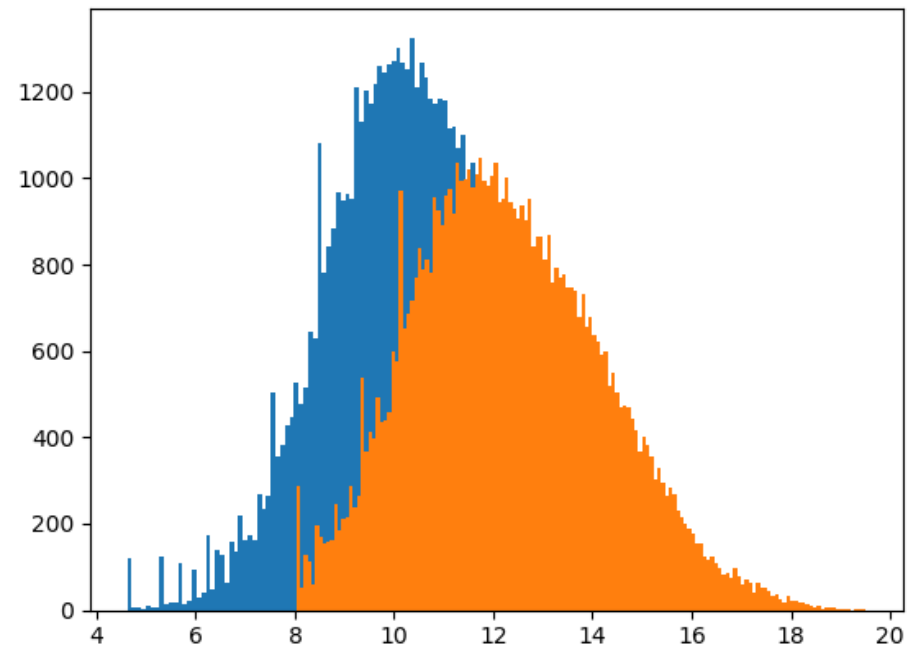

Fig. 3. Empirical distributions of covered and canceled shares for an arbitrarily chosen month (VIRTU). The horizontal axis is plotted in logarithmically transformed coordinates: x = log(1 + #), where # is the share volume.

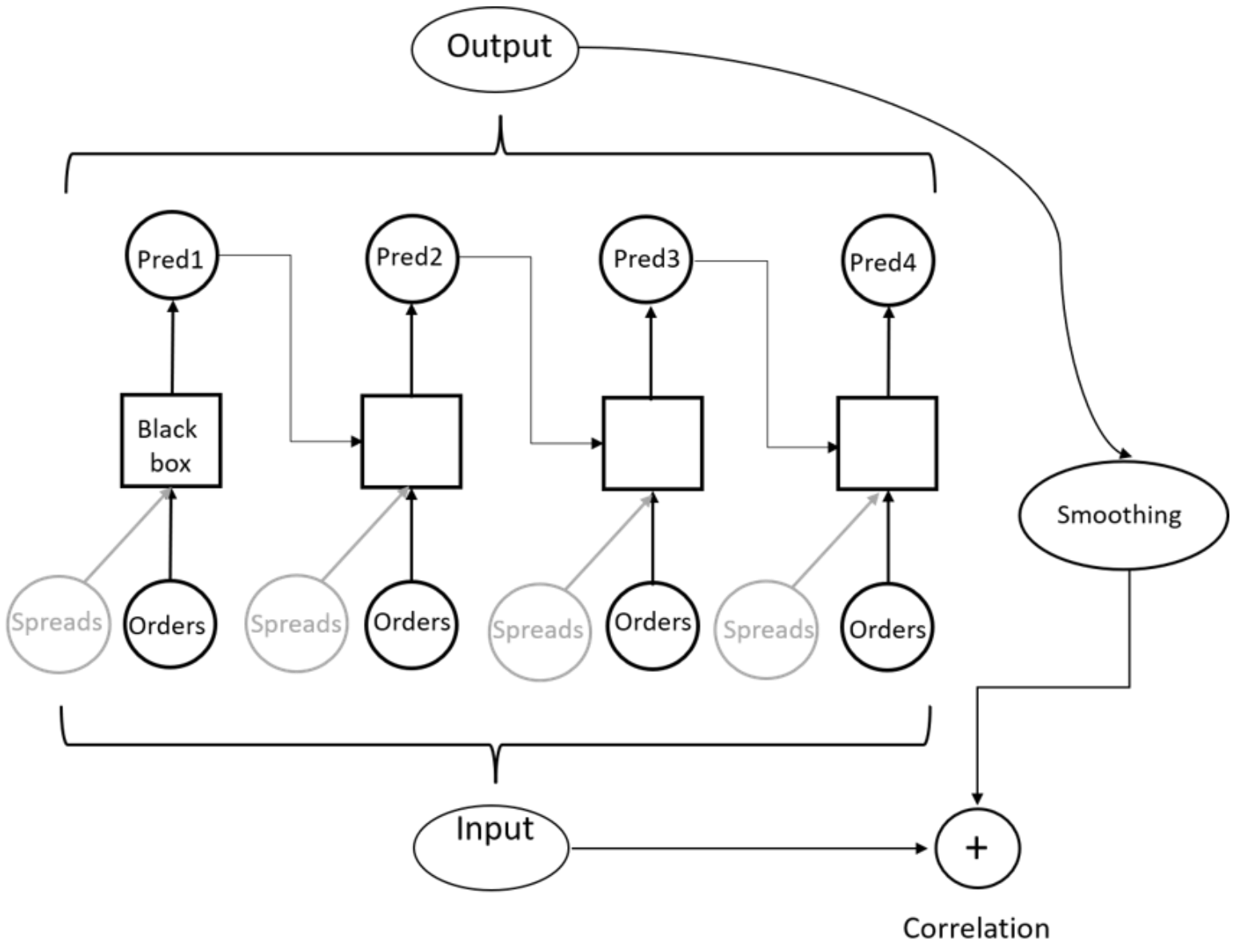

Fig. 4. Architecture of the LSTM (Long Short-Term Memory) network

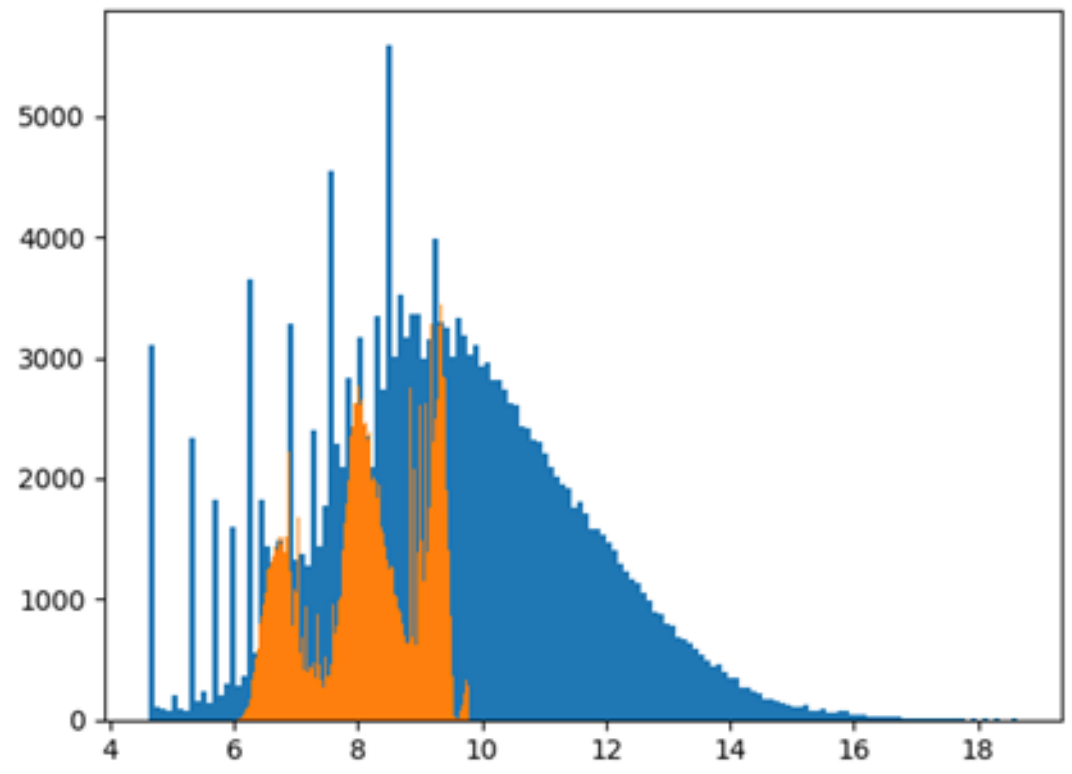

Fig. 4. LSTM prediction of the covered shares distribution in logarithmic coordinates from the spreads (Group 3 distributions). The blue shape is an empirical distribution; the orange shape is a predicted distribution. The prediction quality is very low.

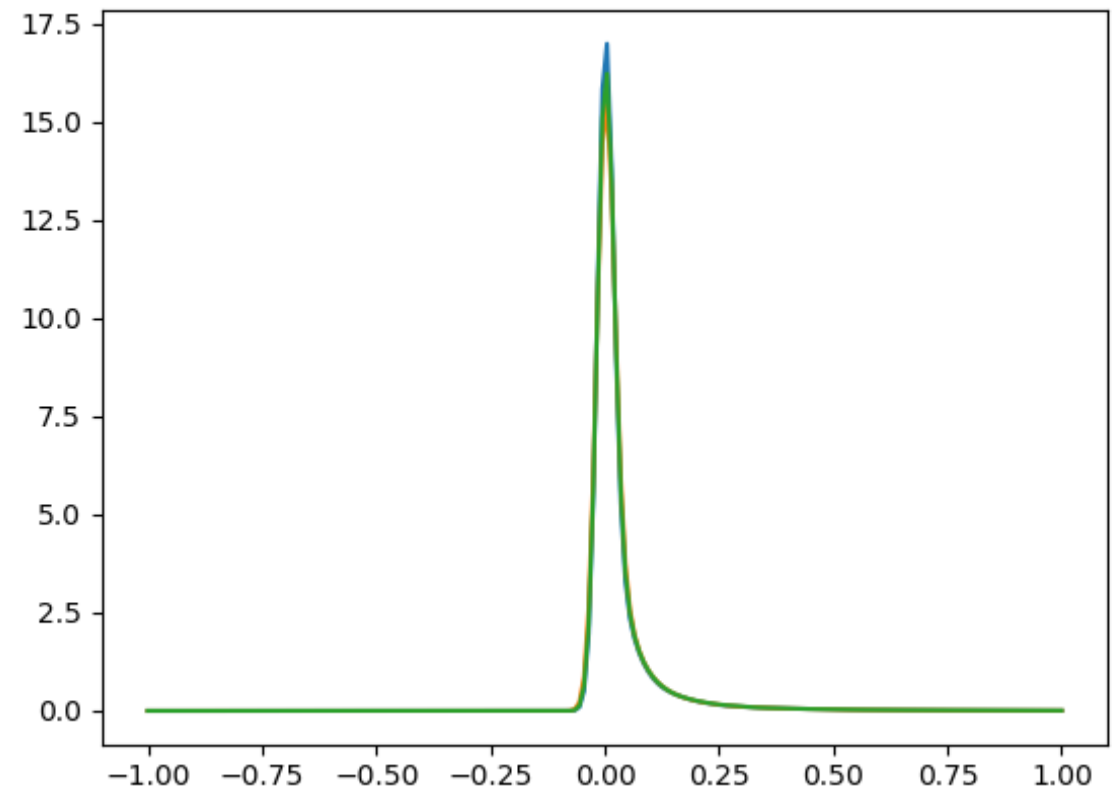

Fig. 5A. Gaussian-smoothed plot of the effective spread (Virtu) for an arbitrary month truncated to the [-1$, 1$] interval. We observe that the probability of an absolute improvement of the effective spread is negligible.

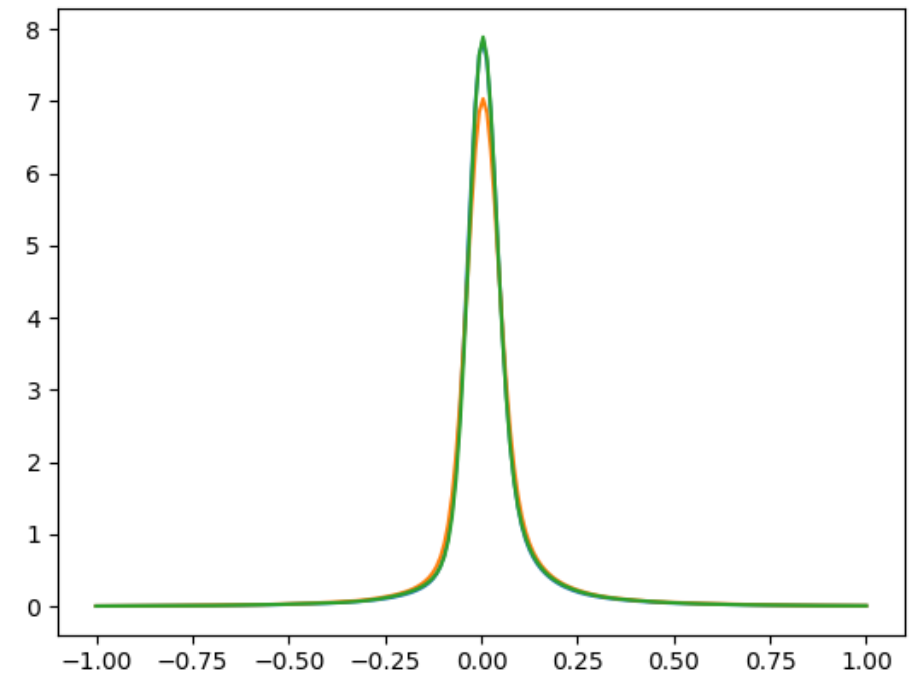

Fig. 5B. Gaussian-smoothed plots of the effective spread (Virtu) for arbitrary months truncated to the [-1$, 1$] interval. We observe that the probability of an absolute improvement of the average spread is almost symmetric with respect to zero.

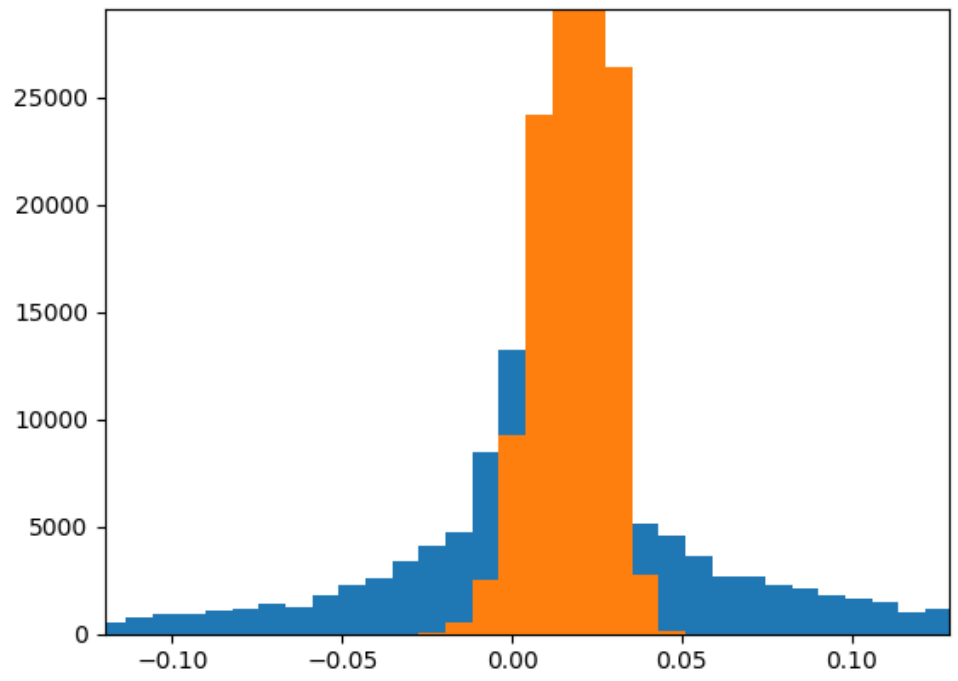

Fig. 6. OLS prediction of the central part [-$0.125, $0.125] of the average spread distribution from the Group 1 distributions of Virtu. The empirical (blue) histogram has fat tails, while the predicted (orange) histogram gravitates toward the center.

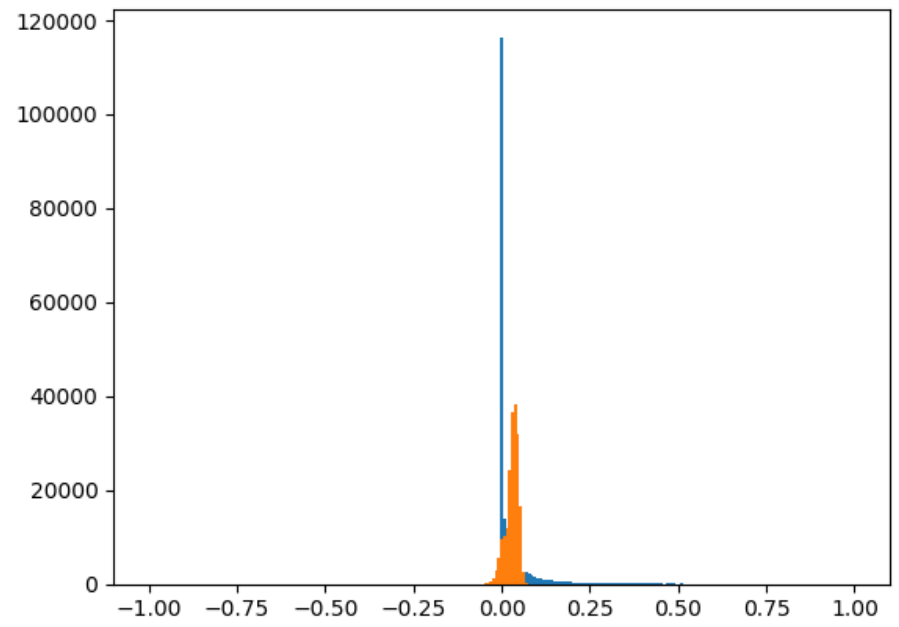

Fig. 7. Same as Fig. 4; restricted to the [-$1, $1] interval.

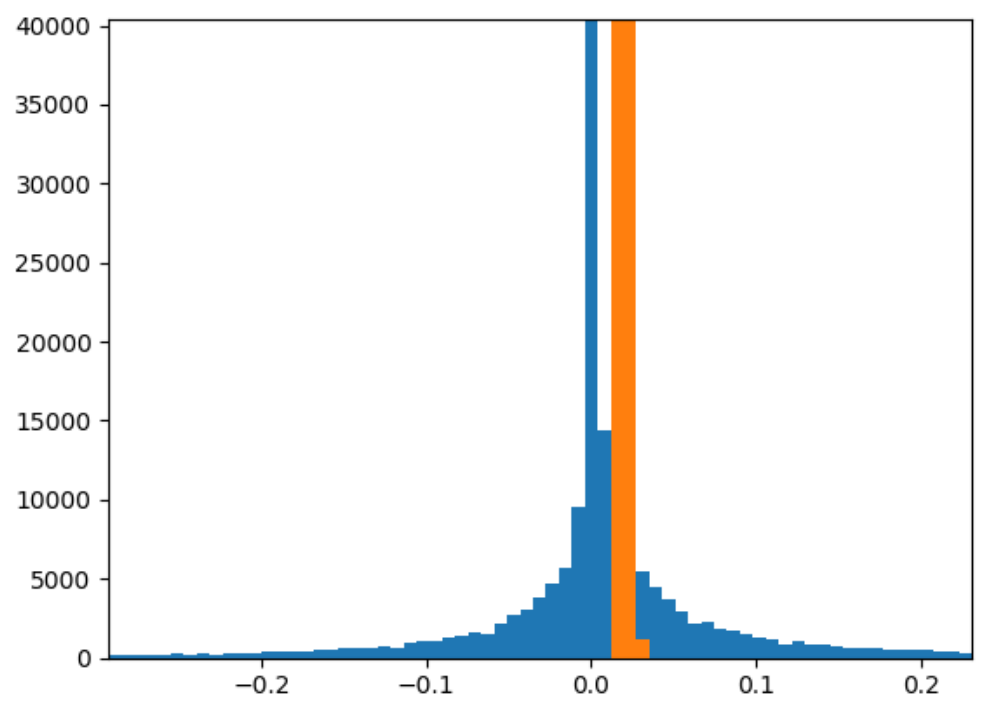

Fig. 8. LSTM prediction of average spreads from the Group 1 variables is even more likely to revert to the mean than the linear regression (compare with Fig. 6).

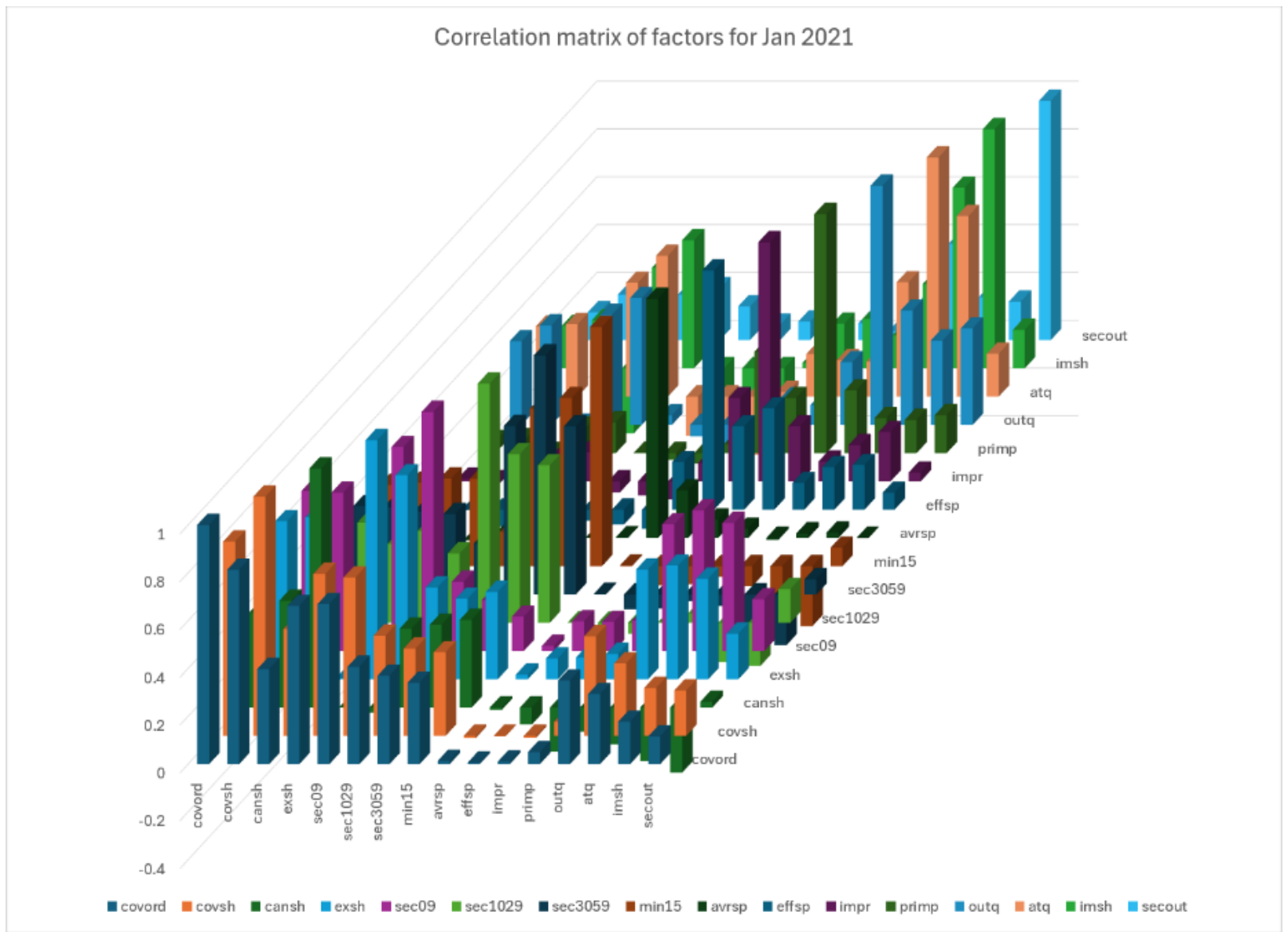

Fig. 9A Correlation matrix for the Virtu data in January 2021

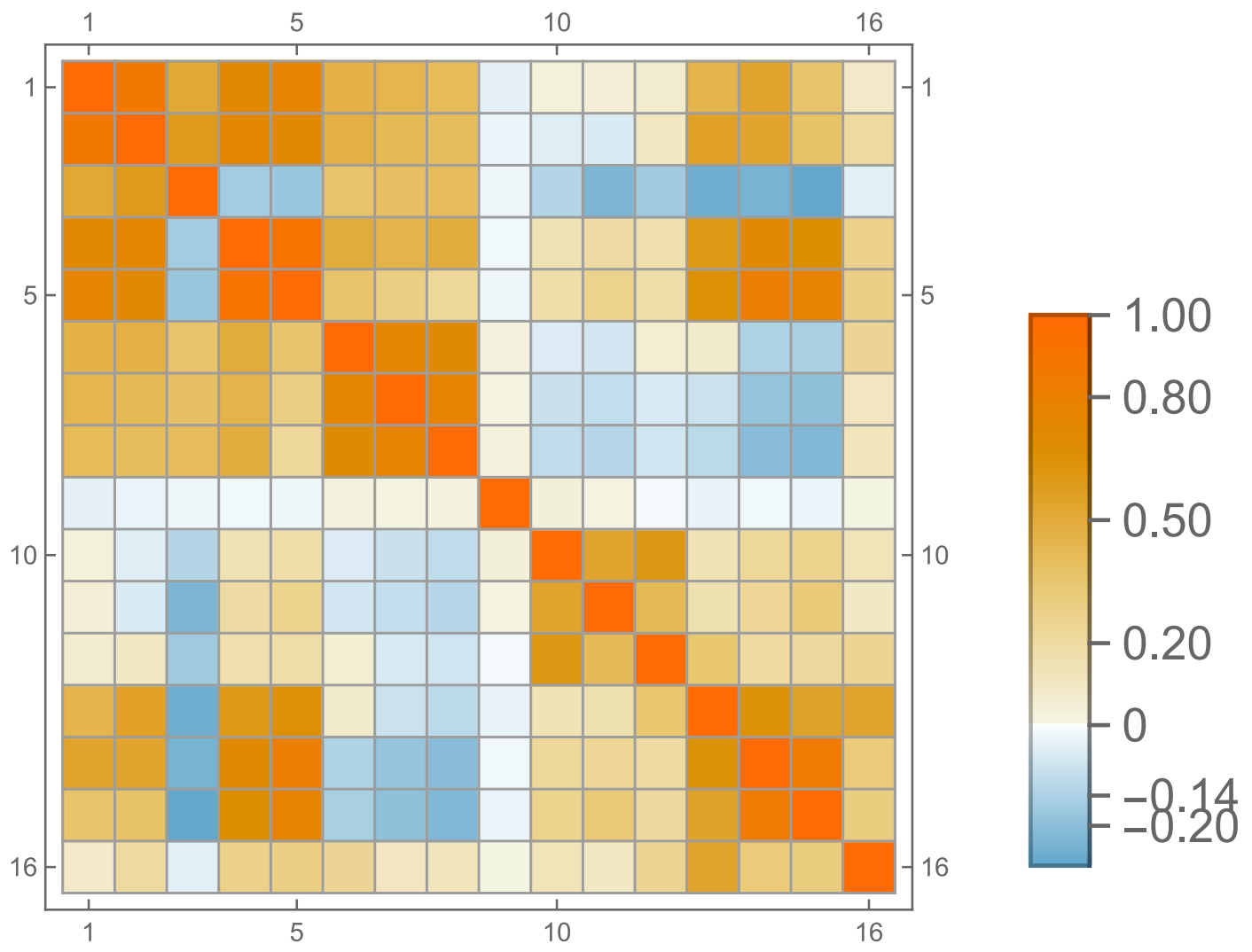

Fig. 9B Sample matrix of correlations of 16 factors of Virtu for May 2022.

A)

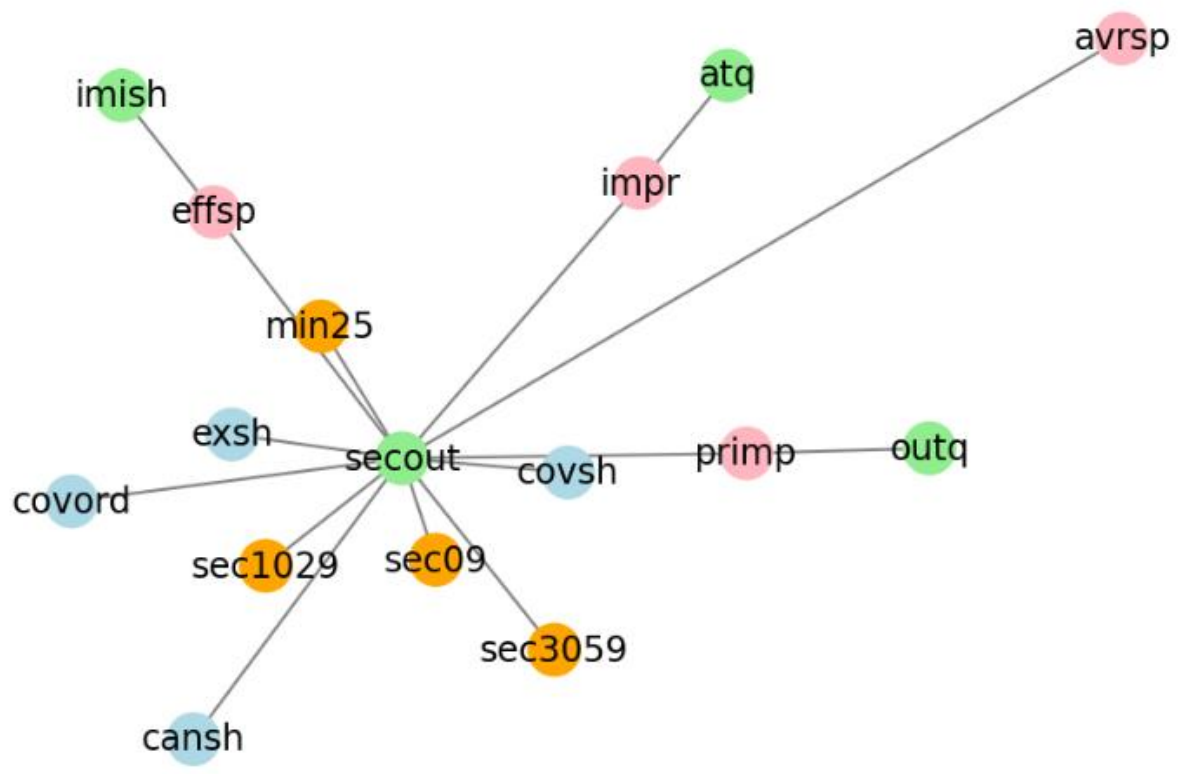

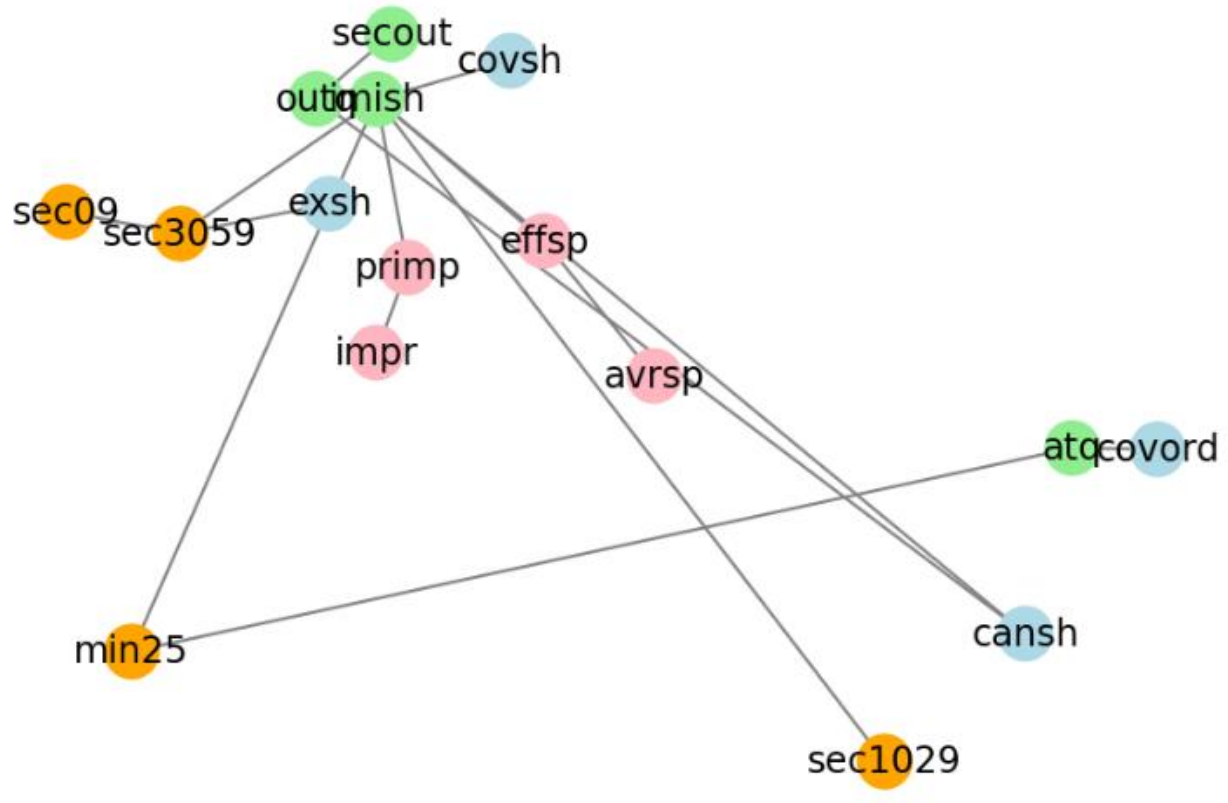

B)

Fig. 10 Plot of the planar graphs of average correlations between factors for the 2021 averages of Virtu (A) and SoHo (B). The color scheme corresponds to groups 1 through 4 in Table 1 (group 1 – blue, group 2 – orange, group 3 – pink, and group 4 – green). The layout is Fruchterman-Reingold. Correlations below ρ=0.15 were omitted for graph visibility.

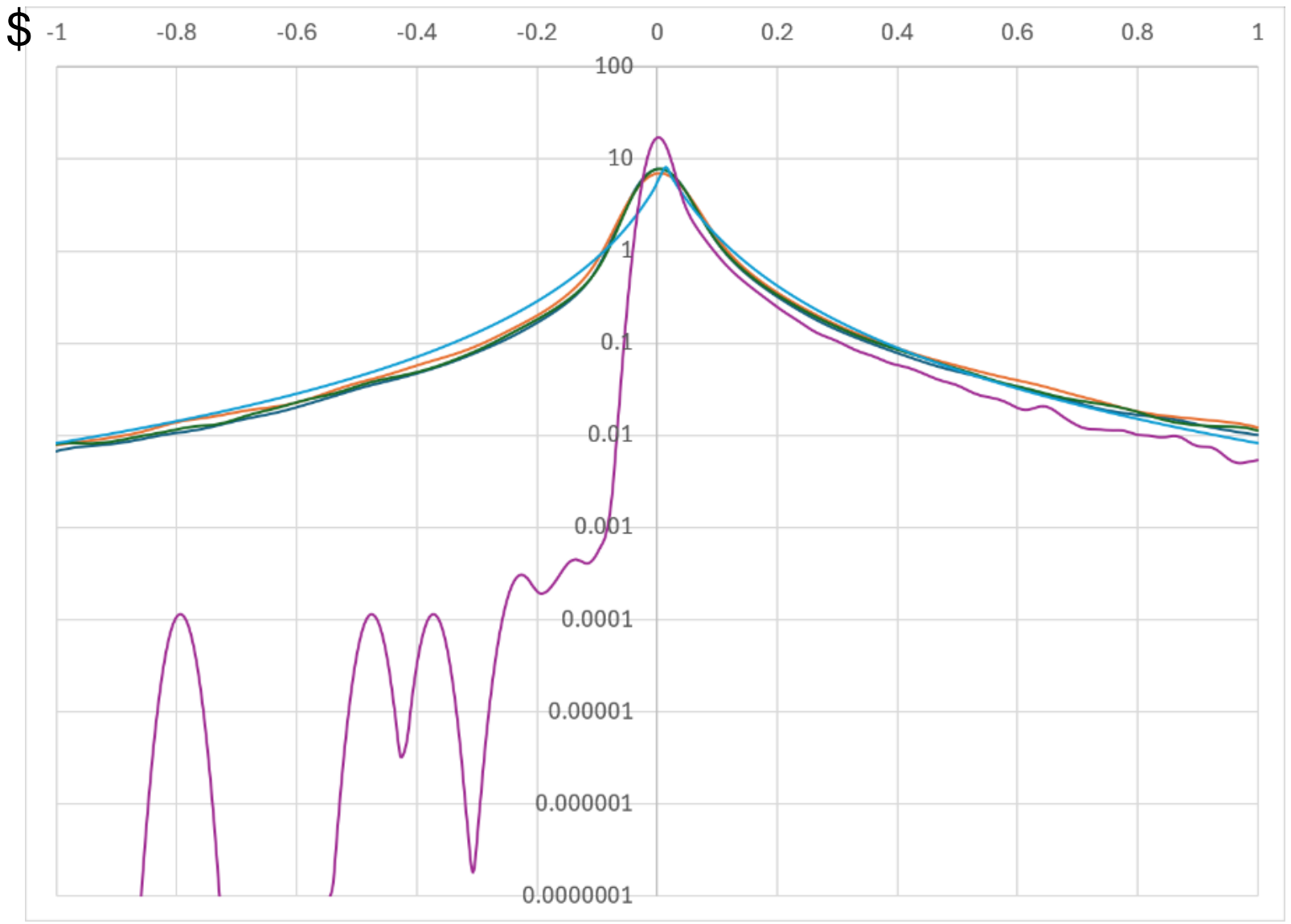

Fig. 11. Logarithmic plots of the average Virtu spreads for 02-04, 2021 (orange, green, dark blue curves for subsequent months, respectively) between -$1 and $1 on the horizontal axis. The average spreads are almost indistinguishable and almost symmetric in the right and left tails (light blue curve with fitted powers $\alpha_1$ = 3.0 and $\alpha_2$ = 2.75). The effective spread for February 2021 is plotted as a purple curve.